\documentstyle[12pt]{article}
\begin{document}
\begin{titlepage}
\title{EXPONENTIALLY LARGE PROBABILITIES IN QUANTUM GRAVITY}

\author{
   O.Yu. Shvedov  \\
{\small{\em Institute for Nuclear Research of the Russian
Academy of Sciences,  }}\\ {\small{\em 60-th October Anniversary Prospect
7a, Moscow 117312, Russia
}}\\ {\small and}\\ 
{\small {\em Sub-faculty of Quantum Statistics and Field Theory,}}\\
{\small{\em Department of Physics, Moscow State University }}\\
{\small{\em Vorobievy gory, Moscow 119899, Russia}} }

\end{titlepage}
\maketitle
\begin{flushright}
 gr-qc/9603039
\end{flushright}
\begin{center}
{\bf \small Abstract}
\end{center}

{\small
The problem of topology change transitions in quantum gravity is investigated
from the Wheeler-de Witt wave function point of view. It is argued that
for all theories allowing wormhole effects the wave function of the 
universe is exponentially large. If the wormhole action is positive, one
can try to overcome this difficulty by redefinition of the inner product,
while for the case of negative wormhole action the more serious problems 
arise.  
}
\newpage

{\bf 1.}
It is known that in quantum mechanics of a particle moving in the external 
potential the semiclassical ground state wave function is exponentially 
small (see, for example, [1]) everywhere, apart from the small region
in the vicinity of the minimum of the potential. The same conclusion about 
exponential smallness is valid for other quantities such as the probability
of the false vacuum decay for the potential with relative minimum
(fig.1a) or the 
instanton shift of levels for the case of double-well-like potential. In 
ordinary quantum field theory models such quantities are also exponentially 
small [2].

I would like to present some examples from the wormhole physics that show us
that these quantitites formally calculated by the semiclassical technique
may occur to be exponentially large in quantum gravity as the parameter of 
the semiclassical expansion tends to zero.

{\bf 2.}
Let us start from consideration of the ground state wave function of the
universe which [3] is the functional of the 3-geometry $g_{ij}({\bf x})$
and of the matter field $\phi({\bf x})$.
Analogously to ref. [4], the wave function can be expressed through the 
functional integral over 4-metrics $g_{\mu\nu}({\bf x},\tau)$ and matter fields
$\phi({\bf x},\tau)$ that start as $\tau\rightarrow -\infty$ from the classical
ground state being the flat space and obey the following boundary condition as 
$\tau=\tau_f$: the values $g_{ij}({\bf x},\tau_f)$ and $\phi({\bf x},\tau_f)$
should coincide with the argument of the wave function under investigation.
Note that in the prescription of [4] the initial 3-geometry was 
considered to be a point, not a flat space. 

The functional integral can be taken by the saddle-point technique
\begin{equation}
\label{*}
\Psi[g_{ij}({\bf x}),\phi({\bf x})] = \int Dg D\phi 
\exp\left(-\frac{1}{\kappa} S[g_{\mu\nu}(\cdot),\phi(\cdot)] \right) \sim
\exp\left(-\frac{1}{\kappa} S\right)
\end{equation}
at small values of the gravitational coupling constant $\kappa$. 
To prove that eq.(\ref{*}) is really a ground state wave function, one can show
that this expression obeys the Wheeler-de Witt equation [3]
and coincides in the 
weak-field approximation with the ground state of the linearized gravity.

If the argument of the wave function is a disconnected 3-geometry, the typical
saddle point being a solution to the euclidean Einstein equations is presented
in fig.2a.
We see that there is evolution from the initial flat 3-geometry (the surface I)
at $\tau=\tau_i=-\infty$ through the singular 3-geometry (dashed line)
to the final disconnected 3-geometry consisting of the large universe
(surface III) and the baby universe (surface II).

{\bf 3.}
Consider the simplest case when there are no matter fields, while the 
interpolating four-geometry is flat. As the gravitational action [3,5]
\begin{equation}\label{act}
S=-\frac{1}{2} \int d^4x \sqrt{g} R + \int d^3{\bf x} \gamma^{1/2} 
K |^{\tau_f}_{-\infty}
\end{equation}
(where
 $\gamma=det g_{ij}$, $R$ is a 4-curvature, $K$ is an external curvature) 
of this solution is equal to $6\pi^2r^2$ (the only contribution comes from the
surface term), the value of the ground state wave function on the disconnected
3-geometry consisting of the flat space and 3-sphere of the radius $r$ is
equal to 
\begin{equation}\label{1}
\exp\left(-\frac{6\pi^2r^2}{\kappa}\right).
\end{equation}
Of course, this quantity is exponentially small. However, we can notice that if 
we increase $\tau_f$, the baby universe will contract, so that the value of
the wave function (\ref{1}) will rapidly increase. This is in contrast with 
quantum mechanical case, where we obtain a suppression of the wave fucntion 
after increasing $\tau_f$. However, in the case of tunneling through the flat
space in quantum gravity one cannot increase $\tau_f$ more and more, because 
the baby universe will contract into a point. This prevents exponential growth
of the wave function. But by adding matter one can allow the wormhole solution
of the type shown in fig.2b. In this case one can increase $\tau_{f}$ arbitrarily
and can therefore expect the value of the wave function to be exponentially 
large.

{\bf 4.}
To confirm this expectation, consider the Giddings-Strominger [6] 
model which is obtained by adding the axionic field $H_{\mu\nu\lambda}=
\partial_{\mu}B_{\nu\lambda}+\partial_{\nu}B_{\lambda\mu}+
\partial_{\lambda}B_{\mu\nu}$, $B_{\mu\nu}=-B_{\nu\mu}$  with the
following additional term of the action:
$$
\Delta S =\frac{\kappa}{12} \int d^4x \sqrt{g} 
H_{\mu\nu\lambda}H^{\mu\nu\lambda}.
$$
By rescaling $H=\tilde{H}/\sqrt{\kappa}$ the integral (\ref{*})
is brought to the saddle-point form. 

Note that we can replace the axionic field by the massless scalar
field; in this case we should consider not coordinate but momentum 
representation for the wave function $\Psi$ (cf. [7,8]): all the
results concerning axionic models will be valid then.

The Giddings-Strominger saddle point is
$$
ds^2=d\xi^2+a^2(\xi)d\Omega_3^2, \tilde{H}_{0ij}=0,
\tilde{H}_{123}=\frac{q}{2\pi^2}\sqrt{det \mu},
$$
where $d\Omega_3^2=
\mu_{11}(d\eta^1)^2+\mu_{22}(d\eta^2)^2+\mu_{33}(d\eta^3)^2$ 
is a metrics on a unit 3-sphere ($\eta_1,\eta_2,\eta_3$ are coordinates
on it), while 
the function $a$ shown in fig.3a is defined up to shift of $\xi$
from the conditions: 
${\mbox sign} (da/d\xi)={\mbox sign} (\xi)$,
$$
|da/d\xi|=\sqrt{1-q^2/(24\pi^4a^4)}.
$$
For given value of the radius of the baby universe $r$, there are two saddle 
points shown in figs.2a,2b.
One of them (fig.2a) comes to the given value of $r$ at once and corresponds
to the negative value of $\xi_f$, another (fig.2b) ''reflects'' from the
turning point $a=(q^2/(24\pi^4))^{1/4}$ and then reaches the value $r$
(the quantity $\xi_f$ is positive).
 Note that at boundary I one has $\tau=-\infty,\xi=-\infty$,
at boundary II : $\tau=\tau_f=\xi_f$, at III : $\tau=\tau_f,\xi=-\infty$.
The action of the euclidean solution consists of two parts: the integral
along the trajectory and the Gibbons-Hawking surface term entering to 
eq.(\ref{act}):
\begin{equation}
\label{acts}
S=\int_{-\infty}^{\xi_f} d\xi [-6\pi^2a(1-a\ddot{a}-\dot{a}^2)
+\frac{q^2}{4\pi^2a^3}] - 6\pi^2a^2(\xi_f)\dot{a}(\xi_f).
\end{equation}
Note that the surface term vanishes at boundaries I and III.

If one replaced these boundaries by the boundary IV (dashed line in fig.2b)
then one would be faced with the infinite contribution of the surface term
and one would be in need of removing it ''by hand'' (one of the prescriptions
is suggested in ref.[9]). However, it follows from the quantum
gravity that one should consider not the boundary IV but the boundaries
I and III. Therefore, there are no infiniteness in the surface term.

Consider  the integral (\ref{acts}) at larges $r$. We see that
the contributions of the saddle points which are calculated 
from eq.(\ref{*}) are:
\begin{equation}\label{2}
\Psi_1\sim\dots \exp\left(-\frac{6\pi^2r^2}{\kappa}\right); 
\Psi_2\sim
\dots i\exp\left(\frac{-\pi q\sqrt{6}/2 + 6\pi^2r^2}{\kappa}\right).
\end{equation}
We see that the second contribution is really exponentially large. 

{\bf 5.} This result 
can be also confirmed by consideration of the minisuperspace Wheeler-de Witt
equation [3] for the function $\Psi[r,H]$ of two variables ; the radius 
of the baby universe and the average value of the axionic field. As there is
an integral of motion -- ''global charge'' -- we can reduce $\Psi[r,H]$ to
$\Psi[r]$, since the dependence on $H$ can be substarcted. The minisuperspace
equation is
\begin{equation}\label{WdW}
\left[ \frac{\kappa^2}{24\pi^2r}\frac{d^2}{dr^2} - 6\pi^2 r +
\frac{q^2}{4\pi^2r^3}\right] \Psi[r] =0,
\end{equation}
If we multiply eq.(\ref{WdW}) by $-r$, we will obtain the Schr\"{o}dinger
equation for the particle moving in the potential shown in fig.1b. The 
problem is how to impose boundary conditions on the minisuperspace wave 
function. Notice that in ordinary quantum mechanics (fig.1a) the radiation
boundary condition (that there are only waves moving out of the classical
vacuunm) is usually imposed. The direct analog of this condition for gravity
is the following: there are no waves moving from the singularity $r=0$
(see fig.1b). If such condition is imposed, one will obtain by the 
semiclassical technique [10,1] the wave function being the sum of 
quantities (\ref{2}), so that the value of $\Psi$ will be exponentially
large. Note that the factor $i$ may play an important role in the 
interpretation of the wormhole as a bounce or as an instanton [11].

Of course, if one considers another boundary conditions (for example, the 
conditions of [12] that there are no singularities as $r\rightarrow 0$)
one will obtain no exponential growth of the wave function. It has been proved
in [13] that even in general case
under certain boundary conditions the wave function cannot be 
exponentially large.

However, it is the second quantity of eq.(\ref{2}) that leads to the 
non-trivial wormhole physics. For example, the diagram shown in fig.4 and
being the foundation of the wormhole calculations [14,15] can be
divided into two subdiagrams, I and II. The contribution of I is proportional
to $\exp(\frac{-\pi q\sqrt{6}/2+6\pi^2 r^2}{\kappa})$, the subdiagram II
is of order $\exp(-6\pi^2 r^2/\kappa)$ because of the surface term, so that
the resulting contribution is $\exp(-S_{WH}/\kappa)$, where $S_{WH}$ is the
whole wormhole action
$$
S_{WH}=\frac{1}{2}\int dx \sqrt{g}R=\int_{-\infty}^{\infty}
d\xi [-6\pi^2a(1-a\ddot{a}-\dot{a}^2+\frac{q^2}{4\pi^2a^3})]=
\frac{\pi q\sqrt{6}}{2}.
$$ 
If we abandon the diagram I because of the  boundary conditions
(or, equivalently, because of the choice of the integration contour),
we should also abandon the contribution of the wormhole shown in fig.4.
Therefore, let us adopt the boundary conditions like shown in fig.1b.

Note also that exponentially large values of $\Psi$ always arise if the
operator $H$ entering to the Wheeler-de Witt equation $H\Psi=0$ has a
discrete spectrum and 0 is not an eigenvalue of $H$.

Since the wave function (\ref{2}) does not belong to $L^2$ (and even to
$S'$), the problem of introducing the inner product and probability 
interpretation of the wave function $\Psi$ arises, since the naive 
interpretating $|\Psi(r)|^2$ as the probability fails.

One of the possible ways to overcome the difficulty is the following
(cf. [4]). Let us define the leading order as $\kappa\rightarrow 0$
of $(\Psi,\Psi)$ as the sum of contributions of saddle points of the
integral 
\begin{equation}\label{i}
\int Dg_{ij} D\phi |\Psi(g_{ij},\phi)|^2.
\end{equation}
If we consider the quantity 
$(\Psi_2,\Psi_2)\sim \int dr \exp(12\pi^2 r^2/\kappa)$, 
there will be no saddle 
points at larges $r$, so there will be no exponentially large contributions
to $(\Psi_1+\Psi_2,\Psi_1+\Psi_2)$, since $(\Psi_1,\Psi_2)$ and 
$(\Psi_1,\Psi_1)$ are exponentially small. Therefore, non-trivial topologies
give rise to the small contribution to eq.(\ref{i}) if $S_{WH}>0$. 
Note also that the latter condition also implies that other wormhole
effects such as shifts of constants of nature [14,15] are small.

{\bf 6.} Is the wormhole action positive for all models? It happens [16] 
that no. Namely, consider the Lavrelashvili-Rubakov-Tinyakov model [17] 
with the action
$$
S=\mu^{-2}\int d^4\tilde{x} \sqrt{g} \left(
-\frac{1}{2}R + \frac{1}{2}g^{\mu\nu}\partial_{\mu}\Phi\partial_{\nu}\Phi
+V(\Phi) + \frac{1}{12}\tilde{H}_{\mu\nu\lambda}\tilde{H}^{\mu\nu\lambda}
\right),
$$
where $\mu$ is a mass parameter, and rescalong $x\rightarrow\mu x=\tilde{x}$ 
is developed to make $\tilde{x}$ dimensionless. Classical equations and
action can be presented as
$$
\left(\frac{d\Phi}{d\xi}\right)^2=-2\frac{d}{d\xi}\left(
\frac{d\ln a}{d\xi}\right)-\frac{2}{a^2}+\frac{q^2}{4\pi^4a^6},
V(\Phi)=-\frac{d^2}{d\xi^2} \ln a + \frac{2}{a^2}-
 3 \left(\frac{d \ln a}{d\xi}\right)^2,
$$
$$
S_{WH}=2\mu^{-2}\int_0^{\infty} d\xi \left[
\frac{q^2}{2\pi^2a^3}-4\pi^2 a + 2\pi^2 \frac{d}{d\xi} 
\left(a^2 \frac{da}{d\xi}\right)
\right].
$$
It is proved in [16] that by varying the potential one can make the 
function $a$ to be equal to the function shown in fig.3b; the contribution
of the region I (where $q^2/(2\pi^2a^3)<4\pi^2 a$) can be made arbitrarily
negative, the contribution of the region II is finite. Therefore, for the
model of [17] $S_{WH}<0$. This means that there are more serious
difficulties in this model, since all wormhole effects formally calculated
by the semiclassical technique will occur to be exponentially large
(of order $10^{10^{38}}$ if $\mu \sim 1 Gev$).

{\bf 7.} Thus, it has been shown that exponentially large values of $\Psi$
arise for all models allowing wormhole effects. For some models, when the
wormhole action is positive, one can try to overcome some of the 
difficulties by redefinition of the inner product. If the wormhole action
is negative, the more serious problems arise, and one should avoid
such models, or suppress topology change by introducing additional
topology coupling [6] (multiplying $n$-wormhole amplitudes by 
$e^{-n\gamma}$ for large positive $\gamma$), or abandon the 
dilute-wormhole-gas approximation being the foundation of the concept of
coupling constants shifts [14,15] for the case of negative wormhole
action.

The author is indebted to G.V.Lavrelashvili,
D.Marolf, Kh.S.Nirov, V.A.Rubakov and P.G.Tinyakov for helpful discussions.
This work was supported in part by ISF, grant \# MKT300.

\begin{center}
{\bf References.}\\
\end{center}
{\bf 1}. L.D.Landau, E.M.Lifshitz. {\it Quantum Mechanics.
Non-Relativistic Theory.} Nauka, Moscow, 1989.\\
{\bf 2}. S.Coleman. {\it Phys. Rev.} {\bf D15} (1977) 2929.\\
{\bf 3}. B.de Witt. {\it Phys. Rev.} {\bf 160} (1967) 1113.\\
{\bf 4}. J.Hartle, S.Hawking. {\it Phys. Rev.} {\bf D28} (1983) 2960.\\
{\bf 5}. G.Gibbons, S.Hawking. {\it Phys. Rev.} {\bf D15} (1977) 2752.\\
{\bf 6}. S.Giddings, A.Strominger.{\it Nucl.Phys.} {\bf B306} (1988) 890.\\
{\bf 7}. K.Lee {\it Phys. Rev. Lett.} {\bf 61} (1988) 263.\\
{\bf 8}. S.Coleman, K.Lee. {\it Nucl.Phys.} {\bf B329} (1990) 387.\\
{\bf 9}. R.Kallosh, A.Linde, D.Linde, L.Susskind.
                {\it Phys. Rev.} {\bf D52} (1995) 912.\\
{\bf 10}. V.P.Maslov. {\it Perturbation Theory and Asymptotic Methods.}
             Moscow, Moscow University press, 1965.\\
{\bf 11}. V.A.Rubakov, O.Yu.Shvedov, in preparation.\\
{\bf 12}. S.Hawking, D.Page. {\it Phys. Rev.} {\bf D42} (1990) 2655.\\
{\bf 13}. D.Marolf, preprint UCSBTH-96-01, gr-qc/9602019.\\
{\bf 14}. S.Coleman. {\it Nucl. Phys.} {\bf B307} (1988) 867.\\
{\bf 15}. S.Giddings, A.Strominger.{\it Nucl. Phys.}{\bf B307} (1988) 854.\\
{\bf 16}. O.Yu.Shvedov, gr-qc/9602049.\\
{\bf 17}. G.V.Lavrelashvili, V.A.Rubakov, P.G.Tinyakov.
       {\it Mod. Phys. Lett.} {\bf A3} (1988) 1231.

{\bf Figure captions.}
\\
{\bf Fig.1}. The potential (solid line) entering to eq.(\ref{WdW}) (fig. 1b)
and the form of the wave function (dashed line) in comparison with the
quantum mechanical case (fig.1a).\\
{\bf Fig.2}. Typical classical euclidean solutions interpolating between
connected (I) and disconnected (II+III) 3-geometries. Dashed lines in 
fig.2a: surfaces $\tau=$const.\\
{\bf Fig.3}. The function $a(\xi)$ for the Giddings-Strominger model
(fig.3a and dashed line in fig.3b) and for the 
Lavrelashvili-Rubakov-Tinyakov model (solid line in fig.3b).\\
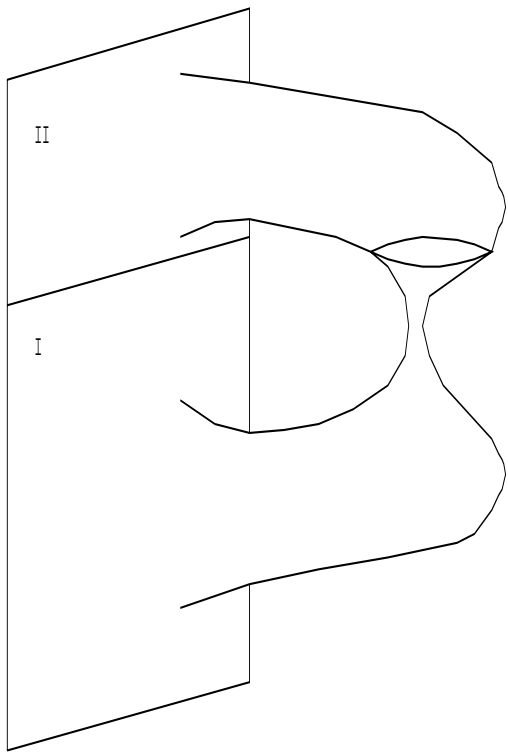
{\bf Fig.4}. The wormhole.
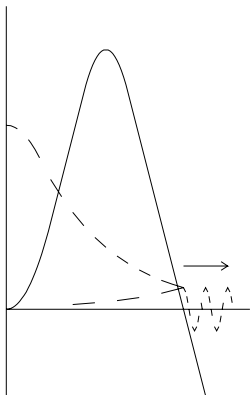
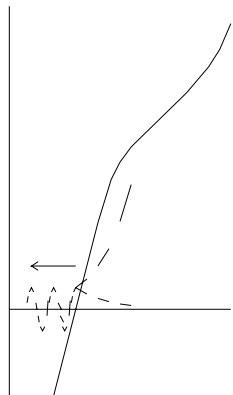
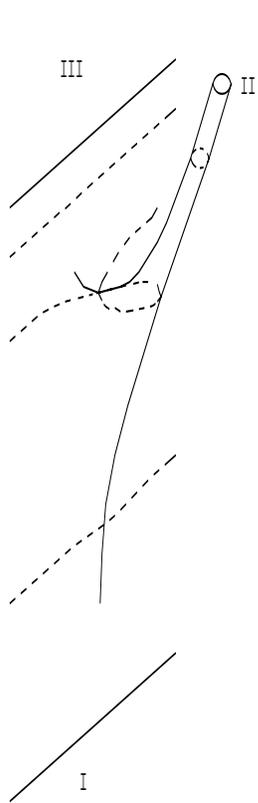
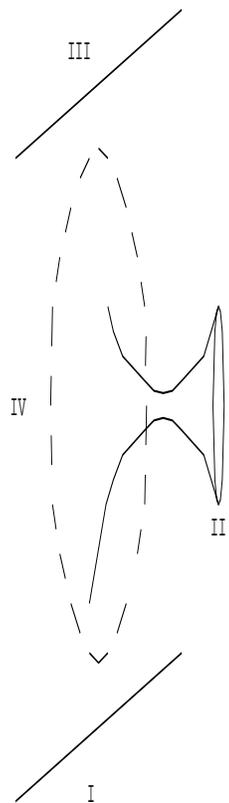
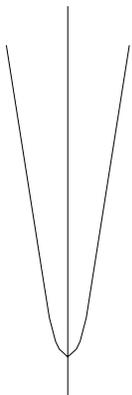
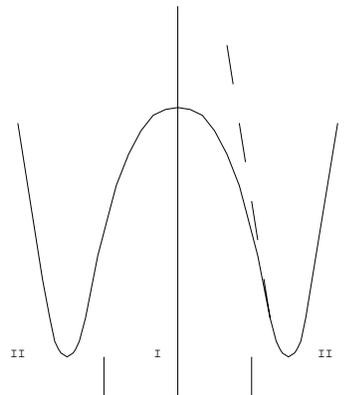
\end{document}